\batchmode
\documentstyle[12pt]{article}
\newcommand{\nc}{\newcommand}
\nc{\renc}{\renewcommand}

\nc{\half}{{\textstyle{1\over2}}}
\nc{\etal}{\mbox{\it et al. }}
\nc{\ie}{{\it i.e.}}
\nc{\eg}{{\it e.g.}}

\renc{\thefootnote}{\arabic{footnote}}
\nc{\capt}[1]{{\bf Figure.} {\small\sl #1}}

\nc{\eqs}[2]{\mbox{Eqs.~(\ref{#1},\,\ref{#2})}}
\nc{\eq}[1]{\mbox{Eq.~(\ref{#1})}}

\nc{\figs}[2]{\mbox{Figs.~(\ref{#1},\,\ref{#2})}}
\nc{\fig}[1]{\mbox{Fig~.(\ref{#1})}}

\nc{\tag}[1]{\label{#1} \marginpar{{\footnotesize #1}}}
\nc{\mtag}[1]{\label{#1} \mbox{\marginpar{{\footnotesize #1}}}}
\renc{\baselinestretch}{1.5}
\newlength{\overeqskip}
\newlength{\undereqskip}
\nc{\be}[1]{\begin{equation} \mbox{$\label{#1}$}}
\nc{\bea}[1]{\begin{eqnarray} \mbox{$\label{#1}$}}
\nc{\Section}[2]{\section{#2}\label{#1}}
\nc{\Bibitem}[1]{\bibitem{#1}}
\nc{\Label}[1]{\label{#1}}

\nc{\eea}{\vspace{\undereqskip}\end{eqnarray}}
\nc{\ee}{\vspace{\undereqskip}\end{equation}}
\nc{\bdm}{\begin{displaymath}}
\nc{\edm}{\end{displaymath}}
\nc{\dpsty}{\displaystyle}
\nc{\bc}{\begin{center}}
\nc{\ec}{\end{center}}
\nc{\ba}{\begin{array}}
\nc{\ea}{\end{array}}
\nc{\bab}{\begin{abstract}}
\nc{\eab}{\end{abstract}}
\nc{\btab}{\begin{tabular}}
\nc{\etab}{\end{tabular}}
\nc{\bit}{\begin{itemize}}
\nc{\eit}{\end{itemize}}
\nc{\ben}{\begin{enumerate}}
\nc{\een}{\end{enumerate}}
\nc{\bfig}{\begin{figure}}
\nc{\efig}{\end{figure}}
\nc{\arreq}{&\!=\!&}
\nc{\arrmi}{&\!-\!&}
\nc{\arrpl}{&\!+\!&}
\nc{\arrap}{&\!\!\!\approx\!\!\!&}
\nc{\non}{\nonumber\\*}
\nc{\align}{\!\!\!\!\!\!\!\!&&}

\def\lsim{\; \raise0.3ex\hbox{$<$\kern-0.75em
      \raise-1.1ex\hbox{$\sim$}}\; }
\def\gsim{\; \raise0.3ex\hbox{$>$\kern-0.75em
      \raise-1.1ex\hbox{$\sim$}}\; }
\nc{\DOT}{\hspace{-0.08in}{\bf .}\hspace{0.1in}}
\nc{\Laada}{\hbox {$\sqcap$ \kern -1em $\sqcup$}}
\nc\loota{{\scriptstyle\sqcap\kern-0.55em\hbox{$\scriptstyle\sqcup$}}}
\nc\Loota{{\sqcap\kern-0.65em\hbox{$\sqcup$}}}
\nc\laada{\Loota}
\nc{\qed}{\hskip 3em \hbox{\BOX} \vskip 2ex}

\nc{\real}{{\rm I \! R}}
\nc{\Z}{{\sf Z \!\!\! Z}}
\nc{\complex}{{\rm C\!\!\! {\sf I}\,\,}}
\def\bigid{\leavevmode\hbox{\small1\kern-3.8pt\normalsize1}}
\def\id{\leavevmode\hbox{\small1\kern-3.3pt\normalsize1}}
\nc{\slask}{\!\!\!/}
\nc{\bis}{{\prime\prime}}
\nc{\pa}{\partial}
\nc{\na}{\nabla}
\nc{\ra}{\rangle}
\nc{\la}{\langle}
\nc{\goto}{\rightarrow}
\nc{\swap}{\leftrightarrow}

\nc{\EE}[1]{ \mbox{$\cdot10^{#1}$} }
\nc{\abs}[1]{\left|#1\right|}
\nc{\at}[2]{\left.#1\right|_{#2}}
\nc{\norm}[1]{\|#1\|}
\nc{\abscut}[2]{\Abs{#1}_{\scriptscriptstyle#2}}
\nc{\vek}[1]{{\rm\bf #1}}
\nc{\integral}[2]{\int\limits_{#1}^{#2}}
\nc{\inv}[1]{\frac{1}{#1}}
\nc{\dd}[2]{{{\partial #1}\over{\partial #2}}}
\nc{\ddd}[2]{{{{\partial}^2 #1}\over{\partial {#2}^2}}}
\nc{\dddd}[3]{{{{\partial}^2 #1}\over
	{\partial #2 \partial #3}}}
\nc{\dder}[2]{{{d #1}\over{d #2}}}
\nc{\ddder}[2]{{{d^2 #1}\over{d {#2}^2}}}
\nc{\dddder}[3]{{d^2 #1}\over
	{d #2 d #3}}
\nc{\dx}[1]{d\,^{#1}x}
\nc{\dy}[1]{d\,^{#1}y}
\nc{\dz}[1]{d\,^{#1}z}
\nc{\dl}[1]{\frac{d\,^{#1}l}{(2\pi)^{#1}}}
\nc{\dk}[1]{\frac{d\,^{#1}k}{(2\pi)^{#1}}}
\nc{\dq}[1]{\frac{d\,^{#1}q}{(2\pi)^{#1}}}

\nc{\cc}{\mbox{$c.c.$ }}
\nc{\hc}{\mbox{$h.c.$ }}
\nc{\cf}{cf.\ }
\nc{\erfc}{{\rm erfc}}
\nc{\Tr}{{\rm Tr\,}}
\nc{\tr}{{\rm tr\,}}
\nc{\pol}{{\rm pol}}
\nc{\sign}{{\rm sign}}
\nc{\bfT}{{\bf T }}

\def\GeV{{\rm\ GeV}}

\nc{\cA}{{\cal A}}
\nc{\cB}{{\cal B}}
\nc{\cD}{{\cal D}}
\nc{\cE}{{\cal E}}
\nc{\cG}{{\cal G}}
\nc{\cH}{{\cal H}}
\nc{\cL}{{\cal L}}
\nc{\cO}{{\cal O}}
\nc{\cT}{{\cal T}}
\nc{\cN}{{\cal N}}
\nc{\rvac}[1]{|{\cal O}#1\rangle}
\nc{\lvac}[1]{\langle{\cal O}#1|}
\nc{\rvacb}[1]{|{\cal O}_\beta #1\rangle}
\nc{\lvacb}[1]{\langle{\cal O}_\beta #1 |}
\nc{\bb}{\bar{\beta}}
\nc{\bt}{\tilde{\beta}}
\nc{\ctH}{\tilde{\cal H}}
\nc{\chH}{\hat{\cal H}}
\nc{\1}{\aa}
\nc{\2}{\"{a}}
\nc{\3}{\"{o}}
\nc{\4}{\AA}
\nc{\5}{\"{A}}
\nc{\6}{\"{O}}
\nc{\al}{\alpha}
\nc{\g}{\gamma}
\nc{\Del}{\Delta}
\nc{\e}{\epsilon}
\nc{\eps}{\epsilon}
\nc{\lam}{\lambda}
\nc{\om}{\omega}
\nc{\Om}{\Omega}
\nc{\ve}{\varepsilon}
\nc{\mn}{{\mu\nu}}
\nc{\k}{\kappa}
\nc{\vp}{\varphi}

\nc{\advp}[3]{{\it  Adv.\ in\ Phys.\ }{{\bf #1} {(#2)} {#3}}}
\nc{\annp}[3]{{\it  Ann.\ Phys.\ (N.Y.)\ }{{\bf #1} {(#2)} {#3}}}
\nc{\apl}[3]{{\it  Appl. Phys. Lett. }{{\bf #1} {(#2)} {#3}}}
\nc{\apj}[3]{{\it  Ap.\ J.\ }{{\bf #1} {(#2)} {#3}}}
\nc{\apjl}[3]{{\it  Ap.\ J.\ Lett.\ }{{\bf #1} {(#2)} {#3}}}
\nc{\app}[3]{{\it Astropart.\ Phys.\ }{{\bf #1} {(#2)} {#3}}}
\nc{\cmp}[3]{{\it  Comm.\ Math.\ Phys.\ }{{ \bf #1} {(#2)} {#3}}}
\nc{\cqg}[3]{{\it  Class.\ Quant.\ Grav.\ }{{\bf #1} {(#2)} {#3}}}
\nc{\epl}[3]{{\it  Europhys.\ Lett.\ }{{\bf #1} {(#2)} {#3}}}
\nc{\ijmp}[3]{{\it Int.\ J.\ Mod.\ Phys.\ }{{\bf #1} {(#2)} {#3}}}
\nc{\ijtp}[3]{{\it Int.\ J.\ Theor.\ Phys.\ }{{\bf #1} {(#2)} {#3}}}
\nc{\jmp}[3]{{\it  J.\ Math.\ Phys.\ }{{ \bf #1} {(#2)} {#3}}}
\nc{\jpa}[3]{{\it  J.\ Phys.\ A\ }{{\bf #1} {(#2)} {#3}}}
\nc{\jpc}[3]{{\it  J.\ Phys.\ C\ }{{\bf #1} {(#2)} {#3}}}
\nc{\jap}[3]{{\it J.\ Appl.\ Phys.\ }{{\bf #1} {(#2)} {#3}}}
\nc{\jpsj}[3]{{\it J.\ Phys.\ Soc.\ Japan\ }{{\bf #1} {(#2)} {#3}}}
\nc{\lmp}[3]{{\it Lett.\ Math.\ Phys.\ }{{\bf #1} {(#2)} {#3}}}
\nc{\mpl}[3]{{\it  Mod.\ Phys.\ Lett.\ }{{\bf #1} {(#2)} {#3}}}
\nc{\ncim}[3]{{\it  Nuov.\ Cim.\ }{{\bf #1} {(#2)} {#3}}}
\nc{\np}[3]{{\it  Nucl.\ Phys.\ }{{\bf #1} {(#2)} {#3}}}
\nc{\npps}[3]{{\it  Nucl.\ Phys.\ Proc.\ Suppl.\ }{{\bf #1} {(#2)} {#3}}}
\nc{\pr}[3]{{\it Phys.\ Rev.\ }{{\bf #1} {(#2)} {#3}}}
\nc{\pra}[3]{{\it  Phys.\ Rev.\ A\ }{{\bf #1} {(#2)} {#3}}}
\nc{\prb}[3]{{\it  Phys.\ Rev.\ B\ }{{{\bf #1} {(#2)} {#3}}}}
\nc{\prc}[3]{{\it  Phys.\ Rev.\ C\ }{{\bf #1} {(#2)} {#3}}}
\nc{\prd}[3]{{\it  Phys.\ Rev.\ D\ }{{\bf #1} {(#2)} {#3}}}
\nc{\prl}[3]{{\it Phys.\ Rev.\ Lett.\ }{{\bf #1} {(#2)} {#3}}}
\nc{\pl}[3]{{\it  Phys.\ Lett.\ }{{\bf #1} {(#2)} {#3}}}
\nc{\prep}[3]{{\it Phys.\ Rep.\ }{{\bf #1} {(#2)} {#3}}}
\nc{\prsl}[3]{{\it Proc.\ R.\ Soc.\ London\ }{{\bf #1} {(#2)} {#3}}}
\nc{\ptp}[3]{{\it  Prog.\ Theor.\ Phys.\ }{{\bf #1} {(#2)} {#3}}}
\nc{\ptps}[3]{{\it  Prog\ Theor.\ Phys.\ suppl.\ }{{\bf #1} {(#2)} {#3}}}
\nc{\physa}[3]{{\it  Physica\ A\ }{{\bf #1} {(#2)} {#3}}}
\nc{\physb}[3]{{\it  Physica\ B\ }{{\bf #1} {(#2)} {#3}}}
\nc{\phys}[3]{{\it Physica\ }{{\bf #1} {(#2)} {#3}}}
\nc{\rmp}[3]{{\it  Rev.\ Mod.\ Phys.\ }{{\bf #1} {(#2)} {#3}}}
\nc{\rpp}[3]{{\it Rep.\ Prog.\ Phys.\ }{{\bf #1} {(#2)} {#3}}}
\nc{\sjnp}[3]{{\it Sov.\ J.\ Nucl.\ Phys.\ }{{\bf #1} {(#2)} {#3}}}
\nc{\spjetp}[3]{{\it Sov.\ Phys.\ JETP\ }{{\bf #1} {(#2)} {#3}}}
\nc{\yf}[3]{{\it Yad.\ Fiz.\ }{{\bf #1} {(#2)} {#3}}}
\nc{\zetp}[3]{{\it Zh.\ Eksp.\ Teor.\ Fiz.\  }{{\bf #1}  {(#2)} {#3}}}
\nc{\zp}[3]{{\it Z.\ Phys.\ }{{\bf #1} {(#2)} {#3}}}
\nc{\ibid}[3]{{\sl ibid.\ }{{\bf #1} {#2} {#3}}}
\nc{\rf}[1]{(\ref{#1})}
\nc{\nn}{\nonumber \\*}
\nc{\bfB}{\bf{B}}
\nc{\bfv}{\bf{v}}
\nc{\bfx}{\bf{x}}
\nc{\bfy}{\bf{y}}
\nc{\vx}{\vec{x}}
\nc{\vy}{\vec{y}}
\nc{\oB}{\overline{B}}
\nc{\oI}{\overline{I}}
\nc{\oR}{\overline{R}}
\nc{\rar}{\rightarrow}
\nc{\ti}{\times}
\nc{\slsh}{\hskip-5pt/}
\nc{\sm}{Standard~Model~}
\nc{\MP}{M_{\rm Pl}}
\nc{\tp}{t_{\rm Pl}}
\nc{\ave}{\bar{E}}

\nc{\eff}{{\rm eff}}
\nc{\kk}{\vek{k}}
\nc{\pp}{{\rm p}}
\nc{\ga}{g_{a\gamma}}
\nc{\vv}{\\}
\nc{\eee}{{\bf E}}
\nc{\bbb}{{\bf B}}
\nc{\qcd}{T_{\rm QCD}}
\nc{\G}{\rm \ G}
\def\vec#1{{\bf #1}}

\def\lae{\;^{<}_{\sim} \;} \def\gae{\; ^{>}_{\sim} \;} 

\def\ell{e^{c}LL}

\begin{document}
{\title{\vskip-2truecm{\hfill {{\small \\
	\hfill \\
	}}\vskip 1truecm}
{\bf  Inflationary Affleck-Dine Scalar Dynamics and Isocurvature Perturbations
    }}
{\author{
{\sc  Kari Enqvist$^{1}$}\\
{\sl\small Department of Physics and Helsinki Institute of Physics,}\\ 
{\sl\small P.O. Box 9,
FIN-00014 University of Helsinki,
Finland}\\
{\sc and}\\
{\sc  John McDonald$^{2}$}\\
{\sl\small Department of Physics and Astronomy, University of Glasgow,
Glasgow G12 8QQ, Scotland}
}
\maketitle
\begin{abstract}
\noindent

We consider the evolution of the Affleck-Dine scalar during D-term and F-term inflation
and solve the combined slow-roll equations of motion.
We show that for a typical case, 
where both the Affleck-Dine scalar and inflaton initially have large values,
in D-term inflation the Affleck-Dine scalar is driven to a fixed value, with only a very
slight dependence on the number of e-foldings.  As a result, there is a definite prediction 
for the ratio of the baryonic isocurvature perturbation to the adiabatic perturbation. In
minimal (d=4) Affleck-Dine baryogenesis the relative isocurvature contribution to the
CMB angular power spectrum amplitude is predicted to be in the range $0.01-0.1$,
which can account for present large-scale structure observations and should be
observable by PLANCK. In a very general case, scale-invariance of the adiabatic
perturbations from the Affleck-Dine scalar imposes a lower bound of about 0.01 for d=4. 
For d=6 the isocurvature perturbation may just be observable, although this is less
certain. We
also consider F-term inflation and show that the magnitude of the baryonic isocurvature
perturbation is fixed by the value of $H$ during inflation. For typical values of $H$
the isocurvature perturbation could be close to present observational limits.

\end{abstract}
\vfil
\footnoterule
{\small $^1$enqvist@pcu.helsinki.fi};
{\small $^2$mcdonald@physics.gla.ac.uk}

\thispagestyle{empty}
\newpage
\setcounter{page}{1}

\section{Introduction}

              With a detailed study of the cosmic microwave background (CMB) planned over
 the coming decade \cite{cmb}, it is important to consider the possible implications for particle physics 
models. The interaction of the particle physics model with the model of inflation may
 generate a CMB which deviates from that expected on the basis of
 inflation alone. We have previously discussed such a case \cite{iso,krev}, the Minimal
 Supersymmetric Standard Model (MSSM) with Affleck-Dine baryogenesis \cite{ad,drt}
 in the context of D-term inflation \cite{dti,lrr}. Affleck-Dine baryogenesis is a very
 natural
and effective candidate for the origin of the baryon asymmetry in SUSY models, 
in particular in the MSSM, where it is the only known candidate in the absence of
 electroweak baryogenesis, for which only a small window of Higgs mass remains \cite{ewb}. In the context of D-term inflation, the Affleck-Dine
 scalar provides a second source of adiabatic perturbations, and requiring that the deviation
 from scale-invariance due to the Affleck-Dine scalar is acceptably small imposes
 an upper bound on the magnitude of the Affleck-Dine scalar, which in turn translates into a
 lower bound on the isocurvature perturbations associated with quantum fluctuations of the
 phase of the Affleck-Dine field \cite{iso}. The spectrum of CMB 
perturbations thus provides a feasible testing bench for Affleck-Dine 
baryogenesis.

        In this paper we wish to consider the dynamical
evolution of the Affleck-Dine (AD) scalar during
 inflation more generally and in more detail. SUSY inflation models are broadly of two
 types, D-term or F-term, depending on the source of the vacuum energy driving inflation \cite{lrr}. 
D-term inflation models have the advantage that the inflaton does not receive order
 $H^{2}$ corrections to its mass squared \cite{dti}, which would prevent slow-rolling
 and produce a highly non-scale invariant 
spectrum of perturbations. Although D-term inflation models have the disadvantage
 that the inflaton field must start at values close to the Planck scale in order to provide
 sufficient inflation \cite{lrr,kmr,lyth}, which requires suppression of Planck-scale
 corrections to the potential, they nevertheless have become the favoured class of
 SUSY inflation models. F-term inflation models 
generically have dangerous order $H^{2}$ corrections to the inflaton mass squared \cite{h2}
(and to the mass squared terms of all other scalars, in particular the AD scalar). 
However, these corrections might be avoided for the inflaton as a result of 
accidental 
cancellations, a special choice of the superpotential and K\"ahler potential
 \cite{sc}, or 
radiative corrections to the inflaton \cite{rc}. We do not, however, expect the
 cancellation to simultaneously apply to any other scalars, and so we expect that in the
 case of F-term inflation, unlike D-term inflation, the AD scalar will have an order
 $H^{2}$ correction to its mass squared term. 
Because of the different mass squared terms during inflation, the dynamics of the AD
 scalar in the two cases will be quite different, with correspondingly different
 consequences for the isocurvature perturbations.

       The paper is organized as follows. In Section 2 we consider the case of D-term
 inflation. We first discuss the slow-rolling dynamics of the
 AD scalar and the inflaton. We then discuss the adiabatic perturbations,
obtaining an upper bound on the magnitude of the AD scalar from scale-invariance
 of the adiabatic perturbations. We next discuss the isocurvature perturbations,
 predicting their magnitude for the case where the AD scalar and inflaton have initially
 large values and more generally obtaining a lower bound from the adiabatic
 perturbation upper bound. In Section 3 we
 consider the case of F-term inflation, showing that when the CMB
 perturbations leave the horizon the AD scalar is likely to be close to the minimum of its potential and that the magnitude of the isocurvature perturbations is then fixed by the 
value of $H$ during inflation. In Section 4 we discuss our conclusions. In an Appendix we briefly review Affleck-Dine baryogenesis. 

\section{D-term Inflation}

\subsection{ Slow-Roll Dynamics of the Affleck-Dine Scalar} 

         D-term inflation \cite{dti} is a form of hybrid inflation \cite{hi}, driven by the
 energy density of a Fayet-Illiopoulos D-term. The inflaton $S$ is coupled to fields
 oppositely charged under a Fayet-Illiopoulos $U(1)_{FI}$ via the superpotential term
\be{d1}   W = \lambda S \psi_{+} \psi_{-}       ~.\ee
The tree-level scalar potential, including the $U(1)_{FI}$ D-term, is then 
\be{d2} V = |\lambda|^{2}  \left( |
\psi_{+} \psi_{-} |^{2} + |S \psi_{+}|^{2} + |S \psi_{-}|^{2}
\right) + \frac{g^{2}}{2} 
\left( |\psi_{+}|^{2} - |\psi_{-}|^{2} + \xi^{2}\right)^{2}   ~,\ee
where $\xi^{2}$ is the FI term and $g$ is the $U(1)_{FI}$ coupling. The global
 minimum of the potential is at 
$S = 0$, $\psi_{+} = 0$, $\psi_{-}= \xi$. However, for $S > S_{crit} \equiv g \xi/\lambda$, the minimum is at $\psi_{+} = \psi_{-} = 0$ and there is a non-zero
 energy density $V_{o} = g^{2} \xi^{4}/2$.  There will be an $S$ potential, however,
 from 1-loop corrections. Thus for $S > S_{c}$ the inflaton potential is given by
 \cite{dti}
\be{e1} V(S) = V_{o} + \frac{g^{4} \xi^{4}}{32 \pi^{2}} ln \left( \frac{S^{2}}{Q^{2}} \right) \;\;\; ; \;\; V_{o} = \frac{g^{2} \xi^{4}}{2}     ~,\ee
where $Q$ is a renormalization scale for the radiative correction. $\xi$
 is fixed by the observed CMB fluctuations \cite{cobe} to be $6.6 \times 10^{15} \GeV$ \cite{lr}.
The total number of e-foldings of inflation, $N$,  remaining at a given value of $S$ 
when the potential is dominated by the inflaton is related to $S$ by
\be{e1a} S = \frac{g N^{1/2}M}{2 \pi}   ~,\ee
where $M = M_{Pl}/\sqrt{8 \pi}$ is
 the mass scale of supergravity (SUGRA) corrections.
The time when the observable CMB perturbations were formed corresponds to $N
 \approx 50$.

      The scalar potential for the AD field $\Phi \equiv \phi e^{i \theta} /\sqrt{2}$ along an
 F- and D-flat direction of dimension $d$ is given by
\be{e2} V(\phi) = \frac{ \lambda^{2} |\Phi|^{2(d-1)} }{M^{2(d-3)}}    ~,\ee
corresponding to a non-renormalizable superpotential term of
 the form $W = \lambda \Phi^{d}/d M^{d-3}$ lifting the flat direction. The coupling $\lambda$ is
 unknown, but if the physical strength of the non-renomalizable interactions is set by
 the SUGRA scale $M$ then we expect that 
$\lambda \approx 1/(d-1)!$ \cite{kmr}. 
In practice, the superpotential term lifting the flat direction is also the B and CP
 violating operator responsible for AD baryogenesis, inducing a baryon asymmetry in
 the 
coherently oscillating $\phi$ condensate (see Appendix). For the case of R-parity
 conserving models, the B violating operators have even dimension, $d = 4,\; 6, ...$. We will refer to the $d=4$ case as minimal AD baryogenesis. 

    For large initial values of $\phi, \ S\sim {\cal O}(M)$, the dynamics
is first dominated by $V(\phi)$. For sufficiently large $\phi$ the
 effective mass squared of the $\phi$ field, $V^{''}(\phi)$, becomes larger than $H^2.$
This occurs once $\phi > \phi_{H}$, where
\be{13a} \phi_{H} = \frac{2^{\frac{(d-1)}{2(d-2)}}}{(6(2d-2)(2d-3))^{\frac{1}{2(d-2)}}} \left(\frac{g}{\lambda}\right)^{\frac{1}{d-2}} \xi^{\frac{2}{d-2}}
 M^{\frac{d-4}{d-2}}      ~.
\ee
 If $\phi_{i} > \phi_{H}$, $\phi$ will initially
 rapidly oscillate in its potential, with an amplitude damped as $\phi \propto a^{-3/d}$, 
where $a$ is the scale factor \cite{turner}. 
However, this period will end before the
 onset of inflaton domination and typically after less than 10 e-foldings of inflation.
The system then enters the regime where both $\phi$ and $S$ are
slowly rolling.

The slow-rolling dynamics of the scalar fields is given by the solution of 
\be{e3} 3 H \dot{\Psi}_{a} = - \frac{\partial V(\Psi_{a})}{\partial \Psi_{a}} \;\;\; ; \;\; 
H = \left(\frac{\sum_{a} V(\Psi_{a})}{3 M^{2}}\right)^{1/2}  ~,\ee
where $\Psi_{a} \equiv S, \; \phi$. 
By taking the ratio of the equations for $\phi$ and $S$ we obtain 
\be{e4}  \frac{\partial \phi}{\partial S} = \frac{16 \pi^{2} (d-1) \lambda^{2}\phi^{(2d-3)} S}{2^{d-2}g^{4} \xi^{4} M^{2(d-3)}}    ~,\ee 
which has the general solution 
\be{e5} \phi = \phi_{i} \left[ 1 + \alpha_{d} \phi_{i}^{2d-4} \left( S_{i}^{2} - S^{2}\right) \right]^{-1/(2d-4)} \;\;\; ; \;\; \alpha_{d}  =  \frac{16 \pi^2(d-2) (d-1) \lambda^{2}}{2^{d-2} M^{2(d-3)} g^4 \xi^4}       ~,\ee 
where $\phi_{i}$ and $S_{i}$ are the initial values at the onset of inflation. 
 We observe two features of this solution. Firstly, since
$S_{i}$ is large compared with the value of $S$ at $N = 50$, we see that 
for sufficiently large $\phi_{i}$ the value of $\phi$ at late times 
is {\it fixed} by $S_{i}$,
\be{e6} \phi \equiv \phi_{*} \approx \left( 
\frac{1}{\alpha_{d}}\right)^{\frac{1}{2d-4}} \frac{1}{S_{i}^{1/(d-2)}}    ~.\ee
This is true if $\phi_{i} > \phi_{*}$, otherwise $\phi$ simply remains at $\phi_{i}$.     
Secondly, we can relate $S_{i}$ to the total number of e-foldings during 
the $V(S)$ dominated period of inflation. In general, for sufficiently large $\phi_{i}$,
 we could have an initial period of $V(\phi)$ dominated inflation. We can show,
 however, that during this period $S$ does not significantly change from $S_{i}$. The
 potential is dominated by $V(\phi)$ once $\phi > \phi_{S}$, where
\be{e7} \phi_{S} = \frac{\sqrt{2} M^{\frac{d-3}{d-1}}}
{\lambda^{\frac{1}{d-1}}}  
\left( \frac{g^{2} \xi^{4}}{2} \right)^{\frac{1}{2(d-1)}}       ~.
\ee 
$\phi_{S}$ is generally less than $\phi_{H}$, therefore $\phi$ will be slow-rolling during $V(S)$ domination.

      From \eq{e5} we find that the condition for 
$S$ to change significantly from $S_{i}$ at a given value of $\phi$ is given by
\be{e8} S_{i}  < \frac{1}{\alpha_{d}^{1/2}} \left(\frac{1}{\phi}\right)^{d-2}   ~.\ee 
Thus the condition for $S$ to change significantly during $V(\phi)$ dominated
 inflation is given by \eq{e8} with $\phi = \phi_{S}$, 
\be{e9} S_{i} < S_{i\;c} \approx \frac{2^{\frac{d-2}{2(d-1)}}}{4 \pi}
 \frac{g^{\frac{d}{d-1}} \xi^{\frac{2}{d-1}} M^{\frac{d-3}{d-1}}}{\lambda^{\frac{1}{d-1}}} 
  ~.\ee
Since $S_{i \; c}$ is small compared with $M$, whereas the value of $S$ required to
 generate 50 e-foldings of inflation, $ S_{50} = g \sqrt{50} M/(2 \pi)$ , is close to
 $M$, it follows that $S_{i}$ ($> S_{50}$) will generally be
 larger than $S_{i\;c}$ and so the inflaton will
 remain at $S_{i}$ until the Universe becomes inflaton dominated. In this case the total
 number of e-foldings of inflation during inflaton domination is given by $N_{S}$,
 where $S_{i} = (g/2 \pi) N_{S}^{1/2} M$. Therefore, if $\phi_{i} > \phi_{*}$, 
$\phi$ at $N \approx 50$ will be given by 
\be{e10} \phi_{*} \approx \left( 
\frac{1}{\alpha_{d}}\right)^{\frac{1}{2d-4}} \left(\frac{2\pi}{g M N_{S}^{1/2}}\right)^{\frac{1}{d-2}}    ~.\ee
The dependence on $N_{S}$ is quite weak; for the case of $d=4$ ($d=6$) Affleck-Dine
 baryogenesis, $\phi_{*} \propto N_{S}^{-1/4}\; (N_{S}^{-1/8})$. Thus if there is
 not an extremely large number of e-foldings of inflation during inflaton domination
 compared with the minimum $N \approx 50$ necessary for the flatness of the
 Universe (i.e. $S$ is not very large compared with $M$), we can essentially fix the value of $\phi_{*}$. In this case we will be able to
 {\it predict} the magnitude of the baryonic isocurvature perturbation.

      It is interesting to speculate on the likely initial values of $\phi$ and $S$. 
The initial value of $S$ is likely to be arbitrary in D-term inflation models, because 
the potential must be very flat even to values of the order of the Planck scale. This is 
because, as noted above, $S_{50}$ is close to the Planck scale, in which case we
 expect Planck scale suppressed superpotential terms to become important. (This is the
 flatness problem of D-term inflation models \cite{kmr,lyth}.) A flat potential can be
 maintained by imposing a symmetry on $S$ (e.g. an R-symmetry \cite{kmr,bbdti}) to
 prevent these dangerous Planck suppresssed terms, so eliminating any potental for
 $S$ beyond the 1-loop
logarithmic term. In this case there is no obvious energy density constraint on the
 initial value of $S$. 
For $V(\phi)$, the energy density rapidly increases as $\phi$ approaches $M$. 
We might then impose a "chaotic inflation" -type initial condition, $V(\phi_{i}) \approx M^{4}$ \cite{ci}. This would give 
\be{e12} \phi_{i} \approx \frac{\sqrt{2}M}{\lambda^{\frac{1}{d-1}}}     ~.\ee
By directly solving the slow-roll equations for $\phi$ and $S$ we can show that the
 total number of e-foldings of inflation is given by 
\be{e11}   N_{T} =  N_{\phi} + N_{S} \approx \frac{1}{4(d-1)} \frac{\phi_{i}^2}{M^{2}} + 
\frac{4 \pi^{2} S_{i}^{2}}{g^{2} M^{2}}     ~,\ee
where $N_{\phi}$ is the number of e-foldings during $V(\phi)$ domination
if $\phi_{i} > \phi_{S}$. From this we see that the $V(S)$ dominated contribution to
 the total number of e-foldings will dominate if 
\be{e13} N_{S} \gae \frac{1}{2 (d-1) \lambda^{2/(d-1)}}    ~.\ee
Since $N_{S} > 50$, this will be satisfied so long as $\lambda$ is not very small (for
 example, if $\lambda \approx 1/(d-1)!$). In this case the value of $\phi$ when the
 CMB perturbations are formed, which in turn 
fixes the magnitude of the isocurvature perturbation, will be determined by the $total$
 number of e-foldings of inflation, $N_{T} \approx N_{S}$.

\subsection{Adiabatic Perturbations from the Affleck-Dine Scalar}

           The potential for the AD scalar is far from flat, and so if the magnitude of the
 AD scalar is large it will cause a large deviation of the adiabatic perturbation from
 scale-invariance. This will impose an upper limit on the magnitude of the AD scalar at
 50 e-foldings. 

      The deviations from scale-invariance are characterized by the spectral index, 
defined so that the density perturbation of present wavenumber $k$ is of the form
 $\delta \rho/\rho \propto 
k^{\frac{n-1}{2}} $ on re-entering the horizon, where \cite{lrr}
\be{e17} n = 1 + 2 \eta - 6 \epsilon    ~.\ee
For the case of a single inflaton $\eta$ and $\epsilon$ are given by the standard
 expressions \cite{lrr,index} 
\be{e18} \eta = M^{2} \frac{V_{S}^{''}}{V_{S}}   ~\ee
and
\be{e19} \epsilon = \frac{M^{2}}{2} \left(\frac{V_{S}^{'}}{V_{S}}\right)^{2}   ~,\ee 
where $V_{S} = V(S)$, $V_{S}^{'} = \partial V/\partial S,\;...$ . 
In order to discuss the influence of the AD scalar, we must generalize these
 expressions to the case of two scalar fields. For a potential of the form 
$V = V(S) + V(\phi)$ we find that, 
\be{e20} \eta = - \frac{M^{2}}{(V_{S}^{'} + V_{\phi}^{'})V}
\left[ V_{S}^{'}V_{S}^{''} + V_{\phi}^{'} V_{\phi}^{''} 
- \frac{2(V_{S}^{'} + V_{\phi}^{'}) (V_{S}^{''}V_{S}^{' \;2} + V_{\phi}^{''} V_{\phi}^{'\;2}) }{(V_{S}^{'\;2} + V_{\phi}^{'\;2})} \right]      ~\ee
and
\be{e21}  \epsilon = \frac{M^{2}}{(V_{S}^{'} + V_{\phi}^{'})V} 
\left[ \frac{(V_{S}^{'} + V_{\phi}^{'}) (V_{S}^{' \;2} +  V_{\phi}^{'\;2}) }{2V}
 \right]    ~.\ee

       For the case of D-term inflation, if $V_{\phi}^{'} < V_{S}^{'}$ and
 $V_{\phi}^{''} < V_{S}^{''}$, we obtain
 the conventional results
\be{e22} \eta = -\frac{1}{2 N}      ~\ee
and 
\be{e23} \epsilon = \frac{g^{2}}{32 \pi^{2} N}    ~.\ee
(The main contribution to scale-dependence therefore comes from $\eta$.) In general, deviation from scale-invariance due to the AD scalar first arises
 when 
$V_{\phi}^{''} > V_{S}^{''}$, with $V_{\phi}^{'} \ll V_{S}^{'}$ and $V_{\phi} \ll V_{S}$ being still satisfied. In this case we can expand $\eta$ to obtain corrections to the
 conventional D-term inflation case, 
\be{e24} \eta \approx M^{2} \frac{V_{S}^{''}}{V_{S}}  - M^{2} 
\frac{V_{\phi}^{'}V_{\phi}^{''}}{V_{S}V_{S}^{'}}  ~.\ee
Thus the deviation of the spectral index from scale invariance due to the AD scalar is 
\be{e25} \Delta n_{\phi} \approx - \frac{2 V_{\phi}^{''}V_{\phi}^{'}M^{2}}{V_{S}V_{S}^{'}}    ~.\ee
Requiring that $|\Delta n_{\phi}| < K$ (present CMB observations imply that $n = 1.2 \pm 0.3$ \cite{cobe}; in the following we will use $K < 0.2$ \cite{lrr}) imposes an upper bound on $\phi$,  
\be{e26}  \phi < \phi_{c} = k_{d} \left( \frac{K}{\sqrt{N}}\right)^{\frac{1}{4 d-7}} 
g^{\frac{5}{4 d-7}} \lambda^{\frac{-4}{4 d-7}} \xi^{\frac{8}{4 d-7}}
M^{\frac{4 d-15}{4 d-7}}    ~,\ee
where 
\be{e27} k_{d} =  \left(\frac{2^{2(d-1)}}{ 128 \pi (d-1)^{2} (2d-3)}
 \right)^{\frac{1}{4 d-7}}    ~.\ee
For the case of minimal $d=4$ Affleck-Dine baryogenesis we obtain
\be{e28} \phi_{c} = 0.53 \left( \frac{K}{\sqrt{N}}\right)^{\frac{1}{9}} 
\left(g^{{5}} \lambda^{-{4}} \xi^{{8}}
M\right)^{\frac 19} \sim 10^{16}\GeV~,\ee
whilst for $d=6$ baryogenesis
\be{e28a} \phi_{c} = 0.77 \left( \frac{K}{\sqrt{N}}\right)^{\frac{1}{17}} 
\left(g^{{5}} \lambda^{-{4}} \xi^{{8}}
M^{9}\right)^{\frac{1}{17}}\sim 10^{17}\GeV   ~.\ee

\subsection{Isocurvature Perturbations from the Affleck-Dine Scalar}

         Isocurvature perturbations of the baryon number arise from the AD scalar 
if the angular direction is effectively massless (i.e. mass small compared with $H$)
 during and after inflation. The resulting 
perturbations will be unsuppressed until the baryon number forms. This in turn requires
 that there are no order $H$ corrections to the SUSY-breaking A-terms. In the effective
 softly broken MSSM at scales $\ll M$, such A-terms can arise only from terms with
 linear couplings of the inflaton superfield to gauge-invariant operators of MSSM
 superfields $\phi_{i}$,
for example, 
\be{e28b} \frac{1}{M} \int d^{2}\theta S W_{i} + h.c.  \sim \frac{F_{S}}{M} W_{i} + h.c.  ~\ee
and 
\be{e28c} \frac{1}{M} \int d^{2}\theta d^{2}\overline{\theta} S \phi_{i}^{\dagger} \phi_{i} + h.c.  \sim \frac{F_{S}^{\dagger}F_{\phi_{i}}\phi_{i}}{M}  + h.c.  ~\ee
In the case of D-term inflation, the inflaton cannot induce an A-term either during
 {\it or} after inflation, since $F_{S} = 0$ in general \cite{kmr}. More generally, if
 there is a symmetry preventing a linear coupling of $S$, then order $H$ A-terms can
 also be eliminated, even in F-term inflation models.   

            The baryon number from
 AD baryogenesis is generated at $H \approx m_{susy} \sim 100 \GeV$ (where
 $m_{susy}$ is the mass scale of the gravity-mediated soft SUSY breaking terms
 \cite{nilles}), when the A-term can introduce B and CP violation into the coherently
 oscillating AD scalar \cite{ad,drt,jrev}. If the phase of the AD scalar relative to the real
 direction (defined by the A-term) is $\theta$, then the baryon number density is (see Appendix) 
\be{b1}  n_{B} \approx m_{susy} \phi_{o}^{2} Sin 2 \theta       ~,\ee
where $\phi_{o}$ is the amplitude of the coherent oscillations at $H \approx
 m_{susy}$. Thus 
\be{b2} \frac{\delta n_{B}}{n_{B}} =  \frac{2 \delta \theta}{Tan(2\theta)}     ~.\ee
$\delta \theta$ is generated as usual by quantum fluctuations of the AD field at horizon
 crossing, 
\be{b3} \delta \theta \approx \frac{H}{2 \pi \phi}    ~,\ee
corresponding to fluctuations of the AD scalar orthogonal to the radial direction. Thus 
\be{b4} \frac{\delta n_{B}}{n_{B}} \approx \frac{H}{\pi \phi Tan (2\theta)}     ~.\ee
The isocurvature perturbation of the CMB is then given by \cite{iso,burns,yana}
\be{e29}  \alpha  = \left| \frac{\delta_{\gamma}^{i}}{\delta_{\gamma}^{a}}\right|  = \frac{\omega}{3} \left( \frac{2 M^{2} V^{'}(S)}{V(S) Tan (2 \theta) \phi}\right)      ~,\ee
where $\delta_{\gamma}^{i}$ is the perturbation in the photon energy density
due to isocurvature perturbations and $\delta_{\gamma}^{a}$ is the perturbation due
 to adiabatic perturbations. For purely baryonic isocurvature perturbations 
\be{e30} \omega = \frac{\Omega_{B}}{\Omega_{m}}    ~,\ee
where $\Omega_{B}$ is the ratio of the energy density in baryons to the critical
 energy density and $\Omega_{m}$ the corresponding ratio for total matter density.
For the case of D-term inflation this gives 
\be{e31} \alpha =   \frac{1}{6 \pi} \frac{g \omega M}{\phi N^{1/2} Tan(2\theta)} 
~,\ee
where $N \approx 50$. 

       Introducing the upper bound on $\phi$ from the requirement that the deviations
 from the spectral
 index due to the AD scalar are acceptably small then gives, for $d=4$, 
\be{e32} \alpha > \alpha_{c} =   \frac{3.3 \omega (g\lambda)^{4/9}}{K^{1/9}
 Tan (2\theta)}
~, \ee
and for $d=6$, 
\be{e32a} \alpha > \alpha_{c} =   \frac{0.18 \omega (g^{3}\lambda)^{4/17}}{K^{1/17}
 Tan (2\theta)}  ~.\ee

       The range of $\Omega_{B}$ allowed by nucleosynthesis is $0.006 \lae
 \Omega_{B}
\lae 0.036$ \cite{sarkar}, where we have taken expansion rate parameter $h$ to be in
 the range $0.6 \lae h \lae 0.87$ \cite{freed}. Thus, for $\Omega_{m} = 0.4$  (in
 keeping with supernova distance measurements \cite{supern}) and $K = 0.2$ we
 obtain for $d = 4$
\be{e33} \alpha_{c} =     (0.06-0.36) \frac{ (g\lambda)^{4/9}}{Tan (2\theta)}
~,\ee
and for $d=6$ 
\be{e33a} \alpha_{c} =     (3.0 \times 10^{-3} - 0.018) \frac{ (g^{3}\lambda)^{4/17}}{Tan (2\theta)}
~.\ee
(The lower limits above should be multiplied by 0.4 for the case $\Omega_{m} = 1$.) 
Thus if, for example, $g \sim \lambda \sim 0.1$ and $Tan (2\theta) \lae 1$, we would
 obtain a lower bound $\alpha \gae 10^{-2}$ for $d = 4$ and $\alpha \gae 10^{-3}$
 for $d=6$. 

It is interesting to note that
    present observations of the CMB combined with large-scale structure from cold dark
 matter (CDM) require that $\alpha \lae 0.1$ \cite{burns,yana}. In particular, using
 COBE 
normalized perturbations combined with the value of $\sigma_{8}$ (the rms of the
 density field on a scale of $8\ {\rm Mpc}$) from X-ray observations of the local cluster
together with the value of the shape parameter ($\Gamma \approx \Omega_{m}h =
 0.25 \pm 0.05$ \cite{pd}) from the galaxy survey (also consistent with recent observations of
 high-redshift supernovae \cite{supern}), 
Kanazawa et al \cite{yana} conclude that $\alpha$ must be less than 0.07 \begin{footnote}{Kanazawa et al define $\alpha$ to be the ratio of the power spectra of the isocurvature to the adiabatic perturbation. This must be multiplied by 16/25 in order to obtain values consistent with our definition of $\alpha$ \cite{eks}.}\end{footnote}. In addition, 
they show that the COBE-normalized 
best fit to large-scale structure ($\sigma_{8}$) in a flat Universe
 with expansion rate parameter
$h \approx 0.7$ (in accordance with recent observations) is given by $\alpha \approx
 0.03 \pm 0.01$. (Large-scale structure
cannot be understood on the basis of CDM with adiabatic perturbations alone.) 
This is exactly in the range expected from minimal $d=4$ AD baryogenesis in the
 context of D-term inflation. Thus isocurvature perturbations from AD baryogenesis
 may already have been observed, although this conclusion very much depends
on accepting the COBE normalization. 

In any case, 
future CMB observations by MAP will be able to probe down to $ \alpha \approx
 0.1$, whilst PLANCK (with CMB polarization measurements) should be able to see
 isocurvature perturbations as small as $0.04$ \cite{eks}. Thus for the case of minimal 
($d=4$) AD baryogenesis, if inflation is D-term then there is a good chance that
 PLANCK will be able to 
observe isocurvature perturbations. For higher dimension AD baryogenesis ($d \geq
 6$) it is less certain, but if $\phi$ is an order of magnitude below the upper
 bound from adiabatic perturbations we could still observe the isocurvature
 perturbations.

        All this assumes that $\phi$ can take any value. This is true if $\phi_{i} <
 \phi_{*}$, in which case $\phi$ remains at its initial value $\phi_{i}$. However, we
 have seen that the dynamics of the AD field during D-term inflation implies that if
 $\phi_{i} > \phi_{*}$ then $\phi$ will equal $\phi_{*}$ at $N \approx 50$. 
In this case we can fix the magnitude of the isocurvature perturbation. 
For $d = 4$, $N \approx 50$ and $\Omega_{m} = 0.4$ the magnitude of the isocurvature perturbation is given by, 
\be{e34} \alpha = \alpha_{*} \approx  (0.17-1.03) \left(\frac{N_{S}}{50}\right)^{1/4}
\frac{(g\lambda)^{1/2}}{Tan(2\theta)}    ~.\ee 
(For $\Omega_{m} = 1$ this should be multiplied by
 0.4.)  For $d=6$ and $\Omega_{m} = 0.4$, 
\be{e34a} \alpha = \alpha_{*} \approx  (4.4 \times 10^{-3}-2.6 \times 10^{-2}) \left(\frac{N_{S}}{50}\right)^{1/8}
\frac{g^{3/4} \lambda^{1/4}}{Tan (2\theta)}    ~.\ee 
If $g, \lambda \gae 0.1$ then for the $d=4$ case we expect $\alpha_{*} \approx
 0.01-0.1$, which is likely to be observable, with the value of the isocurvature
 perturbation being about three times the lower bound expected from the adiabatic perturbation.
 For the $d=6$ case the isocurvature perturbation may just be observable if the baryon
 asymmetry is close to the upper bound imposed by nucleosynthesis and $g$,
 $\lambda$ and $\theta$ take on favourably large and small values respectively.

      It is important that we can fix the isocurvature perturbation to be not much larger than the
 lower bound coming from adiabatic perturbations. This is because there is typically a
 very small range of values of $\phi$ over which the isocurvature 
perturbation is less than the present observational limit, $\alpha \lae 0.1$, but larger
 than the adiabatic perturbation lower bound, $\alpha \gae 0.01$ for $d=4$.
If $\phi$ was more than an order of
 magnitude below its adiabatic upper bound, we would expect to have seen the
 isocurvature perturbation already, and in general there is no reason for the value of
 $\phi$ to be close to the adiabatic upper bound. However, we have shown that the
 case where $S$ and $\phi$ have large initial values ("chaotic inflation" initial conditions)
 provides a natural explanation for a small but potentially observable value of the
 isocurvature perturbation.

\section{F-term Inflation} 

       The results for the case of D-term inflation are based on (i) the absence of order
 $H^{2}$ corrections to the mass squared terms of the AD scalar during inflation and
 (ii) the absence of order $H$ corrections to the A-terms both during and after
 inflation. 
As discussed in the Introduction, models based on F-term inflation must assume that
 the problem of order $H^{2}$ corrections to the 
mass squared of the inflaton has been solved. In this case we can still have 
isocurvature perturbations associated with the Affleck-Dine scalar {\it if} there are no
 order $H$ corrections to the A-terms, which will be the case if there is a symmetry
 forbidding a linear coupling of the inflaton superfield to gauge-invariant operators
 made of MSSM superfields, e.g. a discrete symmetry $S \leftrightarrow -S$ or an R-symmetry.
 However, in the case of F-term inflation we expect in general that the AD scalar will
 have an order $H^{2}$ 
mass squared term during inflation. If this correction were positive in sign, the
 minimum of the potental would be at $\phi = 0$ and the 
AD field would be damped to be exponentially close to $\phi = 0$ by the end of
 inflation, preventing AD baryogenesis. Thus the order $H^2$ correction must be
 negative. This will {\it fix} the 
value of the AD scalar during inflation to be at the $\phi \neq 0$ minimum of its
 potential, which is essentially fixed by $H$ and $d$. This in turn will fix the magnitude
 of the isocurvature perturbation in F-term inflation. 

        During F-term inflation, the potential of the AD scalar is given by 
\be{f1} V_{total}(\phi)  =  -\frac{c H^{2}\phi^{2}}{2} + V(\phi)    ~,\ee
where $V(\phi)$ is the usual potential from the non-renormalizible superpotential term
 and $c \approx 1$.
The minimum is at 
\be{f2} \phi_{m} = \left(\frac{2^{d-2}c}{(d-1)\lambda^{2}}\right)^{1/(2d-4)}
\left(H^{2} M^{2(d-3)}\right)^{1/(2d-4)}     ~.\ee

      Let us first note that if $\phi$ is close to $\phi_{m}$ ($|\delta \phi| \equiv
 |\phi-\phi_{m}| \lae \phi_{m}$) then inflation will damp $\delta \phi$ to be close to
 zero. 
The equation of motion for perturbations around the minimum is
\be{f3} \delta \ddot{\phi} + 3H \delta \dot{\phi} = - k H^{2} \delta \phi \;\; ; \; 
k = (2d-4)c \gae 1  ~.\ee
This has the solution 
\be{f4} \delta \phi = \delta \phi_{o}e^{\alpha H t} \;\;\; ; \;\,  \alpha = \frac{1}{2} 
(-3 + \sqrt{9-4k})  ~.\ee
Thus so long as $Ht \gg1$ i.e. there is a significant number of e-foldings of inflation
 before $N \approx 50$, the AD field will be damped to be exponentially close to the
 minimum of its potential. 

     In general, it is likely that the initial value of $\phi$ will not be close to $\phi_{m}$.
 However, we can show that deviation of the adiabatic perturbation from
 scale-invariance imposes that the value of the potential
 at $N \approx 50$ cannot be very much larger
 than $\phi_{m}$. To see this, suppose that $\phi$ is initially much larger than
 $\phi_{m}$ and consider the contribution of $\phi$ to the spectral index. During
 inflation the invariant $\zeta = \delta \rho/(\rho + p)$ is given by $\delta
 \rho/(\dot{\phi}^{2} + \dot{S}^{2})$. Since $\phi$ will not be slow-rolling 
($V^{''}(\phi) \gg H^{2}$) we must have $\dot{S}^{2} \gg \dot{\phi}^{2}$ 
in order to have a nearly scale-invariant spectrum. We can also assume that $\delta
 \rho$ comes mostly from quantum perturbations for the $S$ field, as the $\phi$ field is 
not effectively massless. Therefore $\zeta \propto \left(V(\phi) +
 V(S)\right)^{3/2}/V^{'}(S)$. The deviation from scale-invariance due to the $\phi$
 field is then 
\be{f5}  \Delta n_{\phi} = -\frac{2}{\xi} \frac{d\xi}{dN} 
= - \frac{3 V^{'}(\phi)}{V(\phi) + V(S)} \frac{\partial \phi}{\partial N}   ~.\ee
For $\phi \gg \phi_{m}$ the $\phi$ field will be rapidly oscillating in its potential and
 the change in the amplitude of $\phi$ over an e-folding due to damping by expansion will be $\partial \phi/\partial N
 \sim -\phi$. Therefore requiring that $|\Delta n_{\phi}| < K$ imposes an upper bound
 on $\phi$,
\be{f6} \phi \lae \left(\frac{Kd}{6(d-1) \lambda^{2}}\right)^{\frac{1}{2(d-1)}} 
\sqrt{2} H^{\frac{1}{d-1}} M^{\frac{d-2}{d-1}}  ~.\ee
Thus 
\be{f7} \frac{\phi}{\phi_{m}} \lae 
\left(\frac{d}{6} \right)^{\frac{1}{2(d-1)}}
\left(\frac{K^{1/(2d-2)}}{c^{1/(2d-4)}} \right)
\left(\frac{ \sqrt{d-1} \lambda M}{H} \right)^{\frac{1}{(d-1)(d-2)}}  ~.\ee
For $d=4$, 
\be{f8} \frac{\phi}{\phi_{m}} \lae 
\frac{0.8}{c^{1/4}}
\left(\frac{\lambda M}{H} \right)^{\frac{1}{6}}  ~,\ee
whilst for $d=6$ 
\be{f9} \frac{\phi}{\phi_{m}} \lae 
\frac{0.9}{c^{1/8}}
\left(\frac{\lambda M}{H} \right)^{\frac{1}{20}}  ~,\ee
where we have used $K = 0.2$.
Therefore for typical values of $H$ during inflation, scale-invariance of the density
 perturbations implies that $\phi$ at $N \approx 50$ cannot be much more than an
 order of magnitude greater than $\phi_{m}$. Since there is no reason for $\phi$ to be
 close to this upper limit when $N \approx 50$, it is most likely that $\phi$ will be close to $\phi_{m}$ when the CBR is formed. 

     Given that $\phi \approx \phi_{m}$, the isocurvature perturbation is given by 
\be{f10} \alpha \approx \frac{2 \omega}{3} \frac{H}{Tan(2 \theta) \delta_{\rho} \phi_{m}} 
 ~,\ee 
where $\delta_{\rho}= 3\delta T/T \approx 3 \times 10^{-5}$ is the value of the CBR
 energy density perturbation \cite{cmb}. 
Therefore given $H$ and $d$, the value of $\phi_{m}$ and so the magnitude of the
 isocurvature perturbation is essentially fixed.
For $d=4$ and $\Omega_{m} = 0.4$ we find
 \be{f11} \alpha = (3.1-18.6) \times 10^{2} \; \frac{\lambda^{1/2}}{c^{1/4} Tan(2
 \theta) }
\left(\frac{H}{M}\right)^{1/2}  ~\ee 
whilst for $d=6$
 \be{f12} \alpha = (2.9-17.4) \times 10^{2} \; \frac{\lambda^{1/4}}{c^{1/8} Tan(2
 \theta) }
\left(\frac{H}{M}\right)^{3/4}  ~.\ee 
If we require that $\alpha \lae 0.1$ in order that the isocurvature perturbation has not been observed at present, 
these impose upper bounds $H/M \lae 10^{-7}/\lambda$ (for $d=4$) and 
$H/M \lae 10^{-5}/\lambda^{1/3}$ (for $d=6$). Thus for typical values of $H$ the 
isocurvature perturbation in the F-term inflation case can be close to present
 observational limits. 

\section{Conclusions} 

     We have considered the dynamics of an Affleck-Dine scalar in the MSSM in the
 context of D- and F-term inflation models and the associated adiabatic and baryonic
 isocurvature perturbations. In the case of D-term inflation, if the AD scalar is initially
 large (as one would expect if the fields obeyed chaotic inflation-like initial conditions)
 then $\phi$ at the time when the CMB goes beyond the horizon will be essentially
 fixed, with a weak dependence on the total number of e-foldings of inflation. In this
 case we can predict the magnitude of the 
isocurvature perturbation. For $d=4$ AD baryogenesis this will be typically in the
 range $\alpha = 0.01-0.1$ and is likely to be observable by PLANCK. This is also
 consistent with the value $\alpha = 0.03 \pm 0.01$ for which a mixed adiabatic and
 isocurvature perturbation spectrum can account for large scale structure observations
 of $\sigma_{8}$, the shape parameter $\Gamma$ and the present expansion rate,
 $h \approx 0.7$ (the latter implies from $\Gamma$ that $\Omega_{M} \approx 0.4$,
consistent with observations of Type Ia supernovae), which cannot be
 understood on the basis of adiabatic perturbations alone. Therefore isocurvature
 fluctuations from D-term inflation/$d=4$ AD baryogenesis may already have been
 indirectly observed.  

   More generally, deviation of the adiabatic perturbation from 
scale-invariance due to the AD scalar imposes an upper bound on the 
magnitude of the AD scalar, which in turn imposes a lower bound on the 
isocurvature perturbation. For $d=4$ AD baryogenesis, the lower bound on 
$\alpha$ is greater than 0.01 for typical values of the unknown parameters, 
again suggesting that the isocurvature perturbation can influence large-scale 
structure formation and is likely to be observable by PLANCK. 
To find precisely the expected limit one should perform a simultaneous fit 
of all the relevant cosmological parameters to the simulated data. One 
should take properly into account the correlation between adiabatic and 
isocurvature perturbations, as well as the
degneracy between isocurvature and tensor perturbations,
which can be resolved by the polarisation data \cite{eks}. We should also
like to point out the AD isocurvature fluctuations are not gaussian,
a fact which can be used to further constrain the amplitudes and hence AD
baryogenesis.

     In the case of F-term inflation, the value of the AD scalar when the CMB goes
 beyond the horizon will most likely be at the minimum of its potential, as determined 
by the negative order $H^{2}$ correction to its mass squared term. Thus if there is an
 isocurvature perturbation (which is possible if there are no order $H$ corrections to
 the A-terms, which simply requires that there is no linear coupling of the $S$
 superfield to MSSM fields), its magnitude will be fixed by $d$ and the value of $H$
 during inflation. For $d=4$ ($d=6$) AD baryogenesis, $H \lae 10^{-6}\;(10^{-4})$ 
is necessary for the isocurvature perturbations to be consistent with current
 observations ($\alpha \lae 0.1$). For reasonable values of $H$ the isocurvature
 perturbations can be large enough to be observable by PLANCK, although, unlike the
 case of D-term inflation, there is no strong reason to expect observable perturbations. 

        We previously discussed the D-term inflation case for $d=6$ AD baryogenesis 
with the formation of late decaying Q-balls of baryon number \cite{bballs}. This
 variant of AD baryogenesis,  "B-ball Baryogenesis" \cite{bballs,bbb,bbdti}, is a natural
 possibility in the MSSM. In this case 
the baryonic isocurvature perturbations of $d=6$ AD baryogenesis are amplified by
 being transferred to dark matter neutralinos via late decay of the B-balls, and are naturally
 in the observable range. So observation of isocurvature perturbations by PLANCK,
 combined with the observation of a deviation of the adiabatic perturbation from scale
 invariance as predicted by D-term inflation, would indicate in the context of
 the MSSM either $d=4$ AD baryogenesis with conventional thermal relic neutralino dark matter
 \cite{jkg} or $d=6$ AD baryogenesis with non-thermal neutralino dark matter
 from late-decaying B-balls \cite{bbb,bbbdm}. 

     Clearly the observation of isocurvature perturbations by PLANCK, together with a
 deviation of the density perturbations from scale-invariance consistent with 
D-term inflation, would have profound implications for both inflation and origin of the
 baryon asymmetry. Indeed, the fact that the expected magnitude of the isocurvature
 perturbations from $d=4$ AD baryogenesis is consistent with present observations of
 large-scale structure may already be indirectly telling us something fundamental about
 the nature of inflation and the baryon asymmetry, which hopefully will be clarified by
 direct observations of the density perturbations by PLANCK.
At the very least, some forms of AD baryogenesis can be ruled out by
the forthcoming CMB observations.

\subsection*{Acknowledgements}   This work has been supported by the
 Academy of Finland under the contract 101-35224, the PPARC (UK) and by the European Union TMR network 
ERBFMRX-CT-970122.

\section*{Appendix. Affleck-Dine Baryogenesis}

         The full scalar potential along a flat direction of the MSSM in the early Universe has the form \cite{drt,h2} 
\be{ap1} V(\Phi) \approx (m_{susy}^{2} - cH^{2})|\Phi|^{2} 
+ \frac{\lambda^{2}|\Phi|^{2(d-1)}
}{M_{p}^{2(d-3)}} + \left( \frac{A_{\lambda} 
\lambda \Phi^{d}}{d M_{p}^{d-3}} + h.c.\right)    ~,\ee
where $m_{susy}$ is the gravity-mediated SUSY breaking mass term, typically of the order of $100 \GeV$. 
In both D- and F-term inflation, once inflation ends and the inflaton begins to 
coherently oscillate about the minimum of its potential the AD scalar will have an
 order $H^2$ correction to its mass squared term. (In D-term inflation this is because
 $F_{S}$ is non-zero when $\dot{S} \neq 0$ \cite{kmr}.) In order to have an unsuppressed value of $\phi$ at $H \approx m_{susy}$,
 the order $H^2$ correction should be negative. (In fact, in D-term inflation models, for $|c|$ less than about 0.5 it is possible to have a positive $H^2$
correction and still generate the observed baryon asymmetry \cite{jad}. Here we will concentrate on the negative $H^2$ correction.) The AD scalar sits at the minimum of its potential until $H \approx m_{susy}$, at which time its mass squared term becomes dominated by the gravity-mediated term and changes sign and the AD scalar beings to coherently oscillate about its new minimum at zero. 
The A-term is dependent upon the phase of the AD field and so can induce B and CP violation in the 
coherently oscillating AD field. In the absence of order H corrections to the A-terms, the initial phase $\theta$ of the AD field (relative to the real direction as defined by the A-term) is random and so typically $\approx 1$. When the AD field starts to oscillate
at $H \approx m_{susy}$, 
the A-term is of the same order of magnitude as the mass squared term, and so 
the A-term will cause the mass of the scalars along the real and imaginary direction to 
differ by $O(m_{susy})$. As a result, these will oscillate with a phase difference $\delta \approx 1$. After a few expansion times, the amplitude of the oscillations will become damped by the expansion of the Universe and the A-term, which is proportional to a large power of $\phi$, 
will become negligible, so fixing the B asymmetry in the AD condensate. The B asymmetry is given by 
\be{ap2} n_{B} = i(\dot{\Phi}^{\dagger}\Phi - \Phi^{\dagger}\dot{\Phi})       ~.\ee
With $\Phi = (\phi_{1} + i \phi_{2})/\sqrt{2}$, where $\phi_{1} = \phi_{o}Cos(\theta) Sin (m_{susy}t)$ and $\phi_{2} = \phi_{o}Sin(\theta) Sin (m_{susy}t + \delta)$, the baryon asymmetry is therefore given by  
\be{ap3} n_{B} \approx  \frac{m_{susy} \phi_{o}^{2}}{2} Sin2 \theta \; Sin \delta   ~.\ee

\end{document}